\documentstyle[11pt,epsfig,twocolumn]{article}
\title{{\bf Remarks on unsolved basic problems
of the Navier--Stokes equations
}}

\author{{\bf Alexander Rauh}
\thanks{Lecture given at 3rd International Summer School/Conference,
LET'S FACE CHAOS through NONLINEAR DYNAMICS, Maribor, Slovenia,
24 June-5 July 1996.}\\
{\small  Fachbereich Physik, Carl von Ossietzky Universit\"at,
D-26111 Oldenburg, Germany }}

\date{}

\begin{document}                                                              

\maketitle

\newcommand{\beq}{\begin{equation}}                                            

\newcommand{\eeq}{\end{equation}}                                              

\newcommand{\beqa}{\begin{eqnarray}}                                            
\newcommand{\eeqa}{\end{eqnarray}}                                              
\newcommand{\fie}{\varphi}                                                    
  
\newcommand{\fieq}{\varphi\varphi}                                            
  
\newcommand{\tet}{\vartheta}                                                  
  
\newcommand{\al}{\alpha}                                                      
  
\newcommand{\de}{\delta}                                                      
  
\newcommand{\la}{\lambda}                                                     
  
\newcommand{\om}{\omega}                                                      
  
\newcommand{\ga}{\gamma}                                                      
  
\newcommand{\na}{\nabla}                                                      
  
\newcommand{\ka}{\kappa}                                                      
  
\newcommand{\pa}{\partial}                                                    
  
\newcommand{\jm}{\jmath}

\newcommand{\bdm}{\begin{displaymath}}                                        
  
\newcommand{\edm}{\end{displaymath}}                                          
\newcommand{\R}{{\rm I}\!{\rm R}}

\begin{abstract}
There is renewed interest in the question of whether the Navier--Stokes
equations (NSE), one of the fundamental models of classical physics and widely
used in engineering applications, are actually self-consistent. After 
recalling
the essential physical assumptions inherent in the NSE,  the notion of  
weak  solutions, possible implications for the energy conservation law,
as well as   existence and 
uniqueness  in the incompressible case are discussed. Emphasis
will be placed on the possibility of finite time singularities
and their consequences for length scales which should be consistent with
the continuum hypothesis.
\end{abstract}

\vspace{0.3cm}

\section{Introduction}
\label{sec1}
As computational fluid dynamics makes progress towards the simulation of 
realistic three-dimensional flows, the validity of the Navier--Stokes equations
(NSE) can be tested in a more and more refined way. To put it from an applied
point of view: Before experiments in wind tunnels are
substituted by computer simulations, one should make sure that the underlying
theory is at least self-consistent. As a matter of fact, after the classical
mathematical work by Leray \cite{bib1}, Hopf \cite{hopf},
Ladyzhenskaya \cite{bib2}, Serrin \cite{serrin},
Temam \cite{bib3}, to refer to important contributions in the field, 
there is renewed interest in the fundamentals  of the NSE,
see for instance the monograph of Doering and Gibbon
\cite{bib4}, or a series of papers by Lions  \cite{bib5} and references
therein.

This contribution
focuses on the question of self-consistency which arises, when one of the
assumptions inherent to the NSE, namely the continuum hypothesis, is
confronted with the length scales emerging from solutions of the 
deterministic NSE. First, the NSE will be briefly derived from
physical principles with due attention paid to the continuum hypothesis.
After recalling  the notion of weak solutions, 
the state of the art of mathematical
existence and uniqueness proofs will be indicated. The  implication of 
weak solutions
upon energy conservation will be discussed. The possibility of finite
time singularities will be related to length scales and thus to the
problem of self-consistency.

\section{Derivation of the NSE}
\label{sec2}
The NSE are based on the conservation of mass and on Newton's second law.
 In addition, the more special assumption of a so--called
Newtonian fluid is adopted, which is justified in a great many cases of
hydrodynamic flows. To formulate the conservation laws it is
customary to pick out a connected cluster of molecules contained in volume 
$V_t$ which is deformed in time and translated according to the local 
velocity ${\bf v}({\bf x},t)$ of
the flow. Time derivatives of corresponding magnitudes are conveniently 
evaluated by means of the Reynolds transport theorem
\beqa
\frac{d}{dt} \int_{V_t}dV\, f({\bf x},t)  &\equiv & \nonumber \\\int_{V_t}dV\,
{\left(\frac{df}{dt} \right)}:= \int_{V_t}dV\, \left[
\frac{\partial f}{\partial t}+\mbox{div}(f {\bf v}) \right]
\label{eq1}
\eeqa
where $f$ is a scalar function. If $\rho ({\bf x},t)$ denotes the
mass density, then conservation of mass, namely
\begin{equation}
\frac{d}{dt} \int_{V_t}dV\, \rho  = 0
\label{eq2}
\end{equation}
gives rise to the continuity equation
\begin{equation}
\frac{d \rho}{dt} \equiv \frac{\partial \rho}{\partial t}+\mbox{div} (\rho
{\bf v}) = 0,
\label{eq3}
\end{equation}
which if $\rho$ = $\rho_0$ is  constant, leads to  the incompressibility 
condition
\begin{equation}
\mbox{div}({\bf v})=0.
\label{eq4}
\end{equation}
Newton's second law implies that any change in momentum is caused by external
forces which in continuum physics are described by the volume force density
${\bf f}$ (e.g.\ gravity) and by a tensorial force 
$\mbox{\boldmath
$\Pi$}$. This tensor reflects the influence of the adjacent fluid on a given
fluid particle. The momentum balance reads
\begin{equation}
\frac{d}{dt} \int_{V_t}dV\, \rho {\bf v} = \int_{V_t}dV\, \rho
{\bf f} + \oint_{\partial V_t}dS\, \mbox{\boldmath $\Pi$} \circ
\hat{{\bf n}} 
\label{eq5}
\end{equation}
where $\hat{{\bf n}}\ dS$ is the oriented surface element of the 
volume $V_t$. It is
convenient to separate in $\mbox{\boldmath $\Pi$}$ an isotropic part, the
pressure $p$, which is present also in the hydrostatic case, from the
so--called stress tensor
$\mbox{\boldmath $T$}$
\begin{equation}
{\Pi}_{ik} = -p {\delta}_{ik} + T_{ik}, \quad \quad
i,k=1,2,3.
\label{eq6}
\end{equation}

The Newtonian fluid assumption now amounts to the following linear relations
between $\mbox{\boldmath $T$}$ and the strain (rate) tensor $\mbox{\boldmath 
$S$}$:
\begin{equation}
T_{ik} = \sum_{m,n=1}^{3} C_{ikmn} S_{mn}
\label{eq7}
\end{equation}
with
\beq
S_{mn} = \frac{1}{2} \left( \frac{\pa v_m}{\pa x_n} +
\frac{\pa v_n}{\pa x_m} \right).
\eeq
The 4th rank tensor
 $\mbox{\boldmath $C$}$ is constant and describes the
effect of viscosity.
 In the isotropic case, $\mbox{\boldmath $C$}$ is of the
form
\begin{equation}
C_{ikmn} =  \nu {\delta}_{ik} {\delta}_{mn}+ \mu 
({\delta}_{im} {\delta}_{kn} + {\delta}_{in} {\delta}_{km})
\label{eq8}
\end{equation}
where $\mu$ and  $\nu$ are macroscopic viscosity parameters. 
In the incompressible
case, $\nu$ drops out, and after making use of mass conservation we can 
write down the momentum balance  as follows
\beqa
\int_{V_t}dV\,  [
\rho_0 \frac{\partial {\bf v}}{\partial t} + \rho_0 \left(
{\bf v} \cdot {\bf\nabla} \right) {\bf v}  +
 \nonumber \\
 \mbox{grad}(p) -
 \mu \Delta {\bf v} - \rho_0 {\bf f} ] = 0.
\label{eq9a}
\eeqa
Here $V_t$ is an arbitrary local space volume. To be sure of the existence
of the above integral, one may adopt the sufficient conditions that the
following fields are locally square integrable
\beq
{\bf v}, \,\, \frac{\pa}{\pa t}{\bf v},\,\,
\frac{\pa}{\pa x_i}{\bf v},\,\,\frac{\pa^2}{\pa x_i \pa x_k}{\bf v},\,\,
\frac{\pa}{\pa x_i}p,\,\, {\bf f}.
\eeq
This can be easily seen with the aid of the Schwarz inequality. For instance,
if $\hat{{\bf x}}_i$ is a  cartesian unit vector,
then we can write
\beqa
|\int_{V_t}dV\, \frac{\partial v_i}{\partial t}|^2 \equiv 
|\int_{V_t}dV\, \frac{\partial {\bf v}\cdot \hat{{\bf x}}_i}{\partial t}|^2
\leq  \nonumber \\
V_t\,  \int_{V_t}dV\, \frac{\pa {\bf v}}{\pa t}
\cdot \frac{\pa {\bf v}}{\pa t}.
\eeqa
As will be discussed later on, there may arise difficulties with the
conservation laws when certain weak conditions on the velocity field 
${\bf v}$ are adopted as is customary in the frame of functional analysis. 
From eq.(10), the following standard NSE in the form of partial 
differential equations are inferred
\beq
\rho_0 \left[ \frac{\partial {\bf v}}{\partial t} + \left(
{\bf v} \cdot \mbox{\boldmath $\nabla$} \right) {\bf v}
\right]  =  -\mbox{grad}(p) + \mu \Delta {\bf v} + 
\rho_0 {\bf f}
\label{eq9}
\eeq
where $p$ is determined through the incompressibility condition 
$\mbox{div}({\bf v})=0$.

\section{Continuum assumptions and length scales}
\label{sec3}
The NSE describe macroscopic physical quantities which constitute mean values
with respect to the underlying atomic degrees of freedom. The density $\rho
({\bf x},t)$ at the space point ${\bf x}$,  for instance,
has to be understood as an average over some volume $\Delta V$ centered at
${\bf x}$. If $\Delta V$ is chosen  too small, a single
measurement of $\rho$ may largely deviate from its mean value due to
molecular fluctuations. An estimate for a
physically reasonable lower bound of $\Delta V$ can be deduced from the mean
thermal density fluctuation $\Delta \rho$ as given in standard textbooks of
thermodynamics \cite{bib6}
\begin{equation}
\frac{\Delta \rho}{\rho} = \sqrt{\frac{kT\kappa}{\Delta V}}
\label{eq10}
\end{equation}
where $k$ is the Boltzmann constant, $T$ the absolute temperature and $\kappa$
the compressibility. If we require the relative fluctuation 
$\Delta \rho / \rho$ to be smaller than, say
$10^{-3}$, at $T=300^{\circ}$ Kelvin, then we find that the diameter $d$ 
of the volume $\Delta V$ should be $d \geq 3 \cdot 10^{-7} \mbox{m}$ 
for air, or $d \geq 10^{-8} \mbox{m}$ for water.

As an implication, if solutions of the deterministic 
NSE turn out to vary on a space scale
much smaller than the above lower bounds, then we are outside of the  validity
domain of these equations.
Here  is the point where the self-consistency problem arises.

In the turbulent regime, length scales decrease with increasing Reynolds
number $R$. As is listed  e.g. in \cite{bib4}, the Kolmogorov length 
$\delta_K$ below which eddies
are destroyed by dissipation, is given by ${\delta}_{K} = L/R^{3/4}$
where $L$ is a typical external length, e.g. the diameter of the
containment. As another example the thickness $\delta_B$ of 
a turbulent
boundary layer scales as $\delta_B \sim L/(R \log R)$. If $L$ = 1 cm, 
then  $\delta_K$ and $\delta_B$ reach the continuum limit at 
$R \approx 10^6$.

\section{Weak solutions and energy balance}
\label{sec4}
Since Leray's pioneering work  \cite{bib1},
one has been looking for generalized solutions ${\bf v}(
{\bf x},t)$ of the incompressible NSE in the space time domain
$\Omega_{\tau} := \Omega \times [0,\tau]$ with the following properties:
\begin{eqnarray} 
{\bf v}({\bf x},t=0) & = & \mbox{\boldmath $\alpha$}({\bf x}),  
\\
{\bf v}|_{\partial \Omega} & = & 0,\\
\mbox{div}({\bf v}) & = & 0.
\end{eqnarray}
The above equations correspond to the initial condition, no-slip
boundary condition and incompressibility, respectively.
To establish weak solutions,  test vector fields $\Phi \in {\bf S}$ are 
introduced  with the following properties
\begin{equation}
{\bf D}:=\{\Phi|\,\,\,\Phi \in D(\Omega); 
\hspace{0.5cm}  \mbox{div}(\Phi)=0 \,\,\, \}
\label{eq14}
\end{equation}
where  $D(\Omega)$ is the Schwartz space ($C^{\infty}$ and compact support
in $\Omega$).
Now ${\bf v} ({\bf x},t)$ is called a weak solution
if it is locally square integrable and if 
the following projections of the NSE and the continuity equation
hold for every $\Phi \in {\bf D}$ 
and for every $C^1$ scalar function $\phi$
with compact support in $\Omega$, respectively \cite{serrin} 
\begin{eqnarray}
\int_{o}^{\tau} dt \int_{\Omega} dV [\Phi_k \frac{\pa v_k}{\pa t}-
v_i v_k \frac{\pa \Phi_k}{\pa x_i}-\nu_0 v_k \Delta \Phi_k- \nonumber \\
 \Phi_k f_k ]  =  0, \\
 \int_{\Omega} dV\, {\bf v}\cdot \mbox{grad}(\phi)  = 0  
\label{eq15}
\end{eqnarray}
where $\nu_0=\mu/\rho_0$ denotes the kinematic viscosity and summation 
convention is adopted.
The pressure term  dropped out in (19) due to the solenoidal 
property of $\Phi$.
A typical theorem reads \cite{bib2}:
\vskip 0.3cm
\noindent
{\bf Theorem}: A unique weak solution exists, at least in the time interval $t
\in [0,{\tau}_1]$ with ${\tau}_1 \leq \tau$, provided the initial velocity
field $\mbox{\boldmath $\alpha$}({\bf x}) \in {W_2}^2$ and
the external force density ${\bf f}$ obeys the condition
\begin{equation}
\int_{0}^{\tau} dt  \int_{\Omega} dV \left [ f^2+ {\left( \frac{\partial f}
{\partial t} \right)}^{2} \right ]^{1/2}  < \infty
\label{eq16}
\end{equation}
where ${W_2}^{2}$ denotes the Sobolev space with the second
space derivatives being  square integrable.
\vskip 0.3cm

As should be noticed, even if the condition (\ref{eq16}) on the external field
${\bf f}$ holds for arbitrarily large $\tau$, uniqueness can be guaranteed 
by the above theorem only for the smaller time interval 
$t \in [0,{\tau}_1]$. While this is typical in space dimension three, 
one has ${\tau}_1=\tau$ in the case of two-dimensional flows.

Which price do we have to pay for accepting weak solutions? To discuss
a possible implication for energy conservation, we recall the notion
of weak and strong convergence of a sequence of real functions $
a^{(1)}, a^{(2)},..a^{(N)},..$. This sequence is called to converge weakly 
against the function $a^*$, if for any square integrable function $g$ 
\beqa
\int_{\Omega}dV\,a^{(N)} a^{(N)} < \infty \hspace{0.5cm} 
\rm{and}\nonumber \\
\lim_{N\rightarrow\infty}\int_{\Omega}dV\,a^{(N)}\,g =
\int_{\Omega}dV\,a^* \,g.
\eeqa
It converges strongly, if
\beq
\lim_{N\rightarrow\infty}\int_{\Omega}dV\,a^{(N)} a^{(N)} 
 = \int_{\Omega}dV\,a^* a^*. 
\eeq
In the case of weak convergence, we have the identity \cite{bib1} 
\beqa
 \lim_{N\rightarrow \infty} \{\int_{\Omega}dV\,(a^{(N)}- a^*)^2-
\int_{\Omega}dV\,a^{(N)} a^{(N)}+
\nonumber\\
\int_{\Omega}dV\,a^* a^*\}=0
\eeqa
which is true because  $\int dV\, a^{(N)} a^*$ converges (weakly) to
$\int dV\, a^* a^*$ and  the two non-converging terms 
$\int_{\Omega}dV\,a^{(N)} a^{(N)}$ cancel each other identically.
As a consequence one has in particular \cite{bib1} 
\beq
\liminf_{N\rightarrow \infty} \int_{\Omega}dV\,a^{(N)} a^{(N)} 
\geq \int_{\Omega}dV\,a^* a^*.
\eeq
Here, the equality sign is guaranteed only in the case of strong
convergence where simultaneously $\liminf=\limsup$. 

To derive the energy balance for a sequence of approximations 
${\bf v}^{(N)}$ which converge weakly against a solution ${\bf v}^{*}$
of (19), we use basis functions $\Phi^{(\nu)} \in {\bf D}$ with the properties 
(18) as
\beq
{\bf v}^{(N)}({\bf x},t):=\sum^N_{\nu=1}c_{(\nu)}(t) \Phi^{(\nu)}({\bf x}),
\hspace{0.4cm} c_{(\nu)}\in {\bf R}.
\eeq
It is convenient to introduce the following abbreviation for the kinetic
energy at time $t$
\beq
 E^{(N)}(t):= \frac{1}{2}
\int_{\Omega}dV\, v^{(N)}_k({\bf x},t)v^{(N)}_k({\bf x},t).
\eeq
$E^*(t)$ denotes the energy corresponding to the weak solution 
\beq
{\bf v}^* \equiv {\bf v}^{(N)}+{\bf r}^{(N)}
\eeq
where ${\bf r}^{(N)}$ is the remainder to the approximate field  
${\bf v}^{(N)}$.
We now insert into (19) the above expression for the weak solution ${\bf v}^*$
together with the test field $\Phi={\bf v}^{(N)} \in {\bf D}$ and obtain
\beqa
E^{(N)}(\tau)\hspace{-.3cm}&-&\hspace{-.3cm}E^{(N)}(0)+
\nu_0 \int^{\tau}_0 \,dt \int_{\Omega}dV\,
\frac{\pa v^{(N)}_k}{\pa x_i}\,
\frac{\pa v^{(N)}_k}{\pa x_i}
\nonumber\\
&&-\int^{\tau}_0 \,dt \int_{\Omega}dV\, v^{(N)}_k f_k=R^{(N)}
\eeqa
with
\beqa
R^{(N)}\hspace{-.3cm}&=&\hspace{-.3cm}
\int^{\tau}_0 \,dt \int_{\Omega}dV\, \Bigl [
-v^{(N)}_k \frac{\pa r^{(N)}_k}{\pa t}
+v^*_i r^{(N)}_k \frac{\pa v^{(N)}_k}{\pa x_i}
\nonumber\\
&&+\nu_0 r^{(N)}_k \Delta\,v^{(N)}_k
+r^{(N)}_k f_k \Bigr ].
\eeqa
Apart from partial integrations, we made use of the incompressibility
condition (20) which implies the relation
\beq
\int_{\Omega}dV\, v^{(N)}_i v^{(N)}_k \frac{\pa v^{(N)}_k}{\pa x_i}=
\frac{1}{2} \int_{\pa \Omega}dS\, (v^{(N)})^2\, \hat{{\bf n}}\cdot 
{\bf v}^{(N)}.
\eeq
The above surface integral  vanishes because ${\bf v}^N \in {\bf D}$.
It should be noticed that eq.(29) holds true for any cutoff $N$; it follows
strictly from the definition (19) of a weak solution; in particular,
no approximate projection scheme was adopted as is common in Galerkin
representations.

In the case of strong solutions with ${\bf v}^* \in W^2_2$ in the 
space time domain $\Omega_{\tau}$,  one can show that $R^{(N)}
\rightarrow 0$ in the limit $N \rightarrow \infty$ so that we would
have the physically plausible energy balance
\beqa
\lefteqn{E^*(\tau)+
\nu_0 \int^{\tau}_0 \,dt \int_{\Omega}dV\,
\frac{\pa v^*_k}{\pa x_i}\,
\frac{\pa v^*_k}{\pa x_i}}&&
\nonumber\\
&=&E^*(0)+\int^{\tau}_0 \,dt \int_{\Omega}dV\, v^*_k f_k,
\eeqa
or in words: the kinetic energy at time $\tau$ plus the energy dissipated
up to $\tau$ equals the initial kinetic energy plus the work done by the 
volume force ${\bf f}$ up to time $\tau$. 

 However, if ${\bf v}^*$ is a weak solution,
then we have only the property of boundedness of the integrals in (29) and
(30),
except for the $f_k$-integrals and the initial energy $E^{(N)}(0)$ which 
converges under the 
assumptions specified in  Theorem A. Making use of the inequality (25)
we can write
\[
\liminf_{N\rightarrow \infty} \left [
E^{(N)}(\tau)   +   
 \nu_0 \int^{\tau}_0 \,dt \int_{\Omega}dV\,
\frac{\pa v^{(N)}_k}{\pa x_i}\,
\frac{\pa v^{(N)}_k}{\pa x_i} \right ] =
\]
\vspace{-.3cm}
\beq
 E^*(\tau)+
\nu_0 \int^{\tau}_0 \,dt \int_{\Omega}dV\,
\frac{\pa v^*_k}{\pa x_i}\,
\frac{\pa v^*_k}{\pa x_i}  +  L_*
\end{equation}
where $L_* \geq 0.$ With  $R_*$ denoting the limes inferior of $R^{(N)}$, the
energy balance  (29) reads in the same limit
\beqa
E^*(\tau)+
\nu_0 \int^{\tau}_0 \,dt \int_{\Omega}dV\,
\frac{\pa v^*_k}{\pa x_i}\,
\frac{\pa v^*_k}{\pa x_i}
\nonumber\\
=E^*(0)+\int^{\tau}_0 \,dt \int_{\Omega}dV\, v^*_k f_k+R_*-L_*.
\eeqa
Thus, in the case of  weak solutions there may be  unphysical sources
or sinks (depending on the sign of $R_*-L_*$)
 of kinetic energy due to the presence of singularities. The latter
are connected  with the space gradients of the velocity field, since
$E^{(N)}(t)$, $t\in (0,\tau)$ can be shown  to converge under
rather general assumptions, see also \cite{rauh}.
If  $R_*-L_*<0$, then the kinetic
energy $E^*(t)$ is smaller than physically expected; this is known
as Leray inequality, see e.g. p. 104 of \cite{bib4}.

\section{Uniqueness and finite time singularities}
\label{sec5}
As already mentioned, one gets square integrable solutions ${\bf v}$ under
rather general assumptions on the external data.  The main basic problem
of the NSE
is related to uniqueness which so far is tied to the existence of the
following time integral, for a recent discussion see  \cite{bib4},
\begin{equation}
I(\tau) := \int_0^{\tau} dt \ \| D {\bf v} \|_{\infty}
\label{eq19}
\end{equation}
with the supremum norm 
\beq
 \|D{\bf v}\|_{\infty}:= \max_{i,k} \max_{{\bf x}\in \Omega}
|\frac{\pa v_k}{\pa x_i}|. 
\eeq 
The origin of this integral will be indicated in the Appendix.
Up to now, in three dimensions the existence of $I(\tau)$ 
has been corroborated only for
finite time intervals $\tau$. 
If $I(\tau)$ exists for arbitrarily large $\tau$,
then both uniqueness and existence of weak solutions can be established for 
arbitrarily large times under quite general conditions.

If $I(\tau)$ exists only up to some time ${\tau}^*$, then $\| D {\bf v}
 \|_{\infty}$ is singular at $t={\tau}^*$ in a way,
that there is at least one space point ${\bf x}_0 \in
\Omega$, where one of the  components $\partial v_i / \partial
x_k$ diverges, for instance as follows
\begin{equation}
\left| \frac{\partial v_i({\bf x},t)}{\partial x_k}
\right|_{{\bf x}={\bf x}_0} \longrightarrow
\frac{{\alpha}^2}{({\tau}^*-t)^{\gamma}}, \hspace{0.3cm}
t<{\tau}^*, \hspace{0.3cm} \gamma \geq 1.
\label{eq20}
\end{equation}
Since for $t$ near $\tau^*$
the behaviour (\ref{eq20}) implies  changes of the velocity
field over arbitrarily small length scales, it is in conflict with the
continuum assumption. The length scales  are then small compared to the
diameter of the  volume $\Delta V$ of a fluid particle  with the consequence
that  microscopic molecular forces  come into play and can no longer be
neglected. In other words we are then out of the validity domain of the
deterministic NSE and we would have to consider stochastic forces in
addition to the deterministic external forces. It is therefore not yet
settled, whether the phenomenon of hydrodynamic turbulence is a manifestation
of deterministic chaos alone.

As should be noted, the problem of   finite time singularities 
cannot be overcome by some  averaging recipe, because
the existence of $I(\tau)$ is connected
 to the uniqueness of solutions as a sufficient condition, and it may 
turn out to be also necessary.

Similarly, in the case of compressible flows 
finite time singularities could not  be excluded so far \cite{bib5}. 
The proof or disproof of the
existence of finite time singularities constitute one of the basic unsolved 
problems in the analysis of the NSE. 
In the inviscid case of the Euler equation, there is a general argument
for possible finite time singularities, see for instance Frisch 
\cite{frisch}.  From a direct numerical simulation of the Euler equations,
 Grauer and Sideris \cite{bib7} recently reported  on evidence for 
a singularity of the type as given in   (\ref{eq20}) with $\gamma=1$.

\vspace{1cm}
{\bf Acknowledgement}
The author is indepted to M. Boudourides for making him aware of 
J. Serrin's contribution to the field. He is also thankful to A. Spille
for a critical reading of the manuscript.

\vspace{1cm}
\noindent
{\Large \bf Appendix}

\vspace{0.3cm}
In the following it is  sketched  how the integral $I(\tau)$ shows up
in uniqueness proofs, 
see \cite{bib4}. At variance with \cite{bib4} we do not adopt 
periodic boundary conditions. 
Let us assume there are two different solutions ${\bf v}$ and ${\bf v'}$
of the NSE (13).  Then we define ${\bf u}:={\bf v}-{\bf v'}$ and obtain
after subtracting the NSE for  ${\bf v}$ and ${\bf v'}$
\beq
\frac{\pa u_k}{\pa t}-u_i\frac{\pa u_k}{\pa x_i} +
v_i\frac{\pa u_k}{\pa x_i}+u_i\frac{\pa v_k}{\pa x_i} = 
-\frac{\pa (p-p')}{\pa x_k}+\nu_0 \Delta\, u_k. 
\eeq 
When this equation is scalarly multiplied by ${\bf u}$ and integrated over
the volume $\Omega$, then, apart from the pressure term,
the second and third terms of the left hand side
vanish by the same argument used before in (31). With the
abbreviation 
\beq
\|{\bf u}\|^2=\int_{\Omega}dV\, {\bf u}\cdot {\bf u} 
\eeq
we can write
\[
\frac{1}{2} \frac{d }{d t} \|{\bf u}\|^2  =  A+B; 
\]
\beq
A:=-\nu_0 \int_{\Omega}dV\, \frac{\pa u_k}{\pa x_i}\,\frac{\pa u_k}{\pa x_i};
\,\,\,
B:  =  -\int_{\Omega}dV\, u_k\frac{\pa v_k}{\pa x_i}u_i.
\eeq
Now the viscosity term is estimated by means of the Poincar\'{e} inequality
\cite{joseph}
\beq
-A\equiv |A| \geq \frac{2}{l^2}\, \|{\bf u}\|^2
\eeq
where $l$ denotes the smallest distance between two parallel planes
which just contain $\Omega$. The  $B$ term is estimated by using
 the definition (36) and the Schwarz inequality
as  follows
\[
|B|= |\int_{\Omega}dV\, u_k\frac{\pa v_k}{\pa x_i}u_i| \leq
\|D{\bf v}\|_{\infty}\sum^3_{k,i=1}\int_{\Omega}dV\, |u_k u_i|
\]
\vspace{-.3cm}
\beq
\leq 9 \|D{\bf v}\|_{\infty} \|{\bf u}\|^2.
\eeq
One  arrives at the ordinary differential inequality
\beq
\frac{1}{2} \frac{d}{d t} \|{\bf u}\|^2 \leq \left[-\frac{2 \nu_0}{l^2}+
9\,\|D{\bf v}\|_{\infty} \right] \, \|{\bf u}\|^2,  
\eeq
which by Gronwall's lemma can be integrated to the final inequality
\beq
\|{\bf u}(t)\|^2\leq \|{\bf u}(0)\|^2\exp\left [-\frac{4 \nu_0}{l^2} t+
18 I(t)\right ].
\eeq
This result tells that, since the two supposed solutions ${\bf v}, {\bf v'}$
possess the same initial conditions and therefore ${\bf u}(0)=0$, we have
${\bf u}(t)=0$ for times $t \in (0,\tau)$ for which  $I(t)$ exists.
This conclusion  holds true also in the inviscid
limit $\nu_0 \rightarrow 0$.

\end{document}